# Quantum-interference metrology of dissipative Kerr solitons

Yun-Ru Fan,[1,2,3] Yong Geng,[4] Ji Liu,[1,3] Yong-Jun Huang,[4] Hai-Zhi Song,[1,5] Hao Li,[6] Li-Xing You,[6] Heng Zhou,[4,†] Kun Qiu,[4] Kai Guo,[7,#] Guang-Can Guo,[1,2,3,8] and Qiang Zhou[1,2,3,8,*]

[1]Institute of Fundamental and Frontier Sciences, University of Electronic Science and Technology of China, Chengdu 611731, China

[2]Center for Quantum Internet, Tianfu Jiangxi Laboratory, Chengdu 641419, China

[3]Key Laboratory of Quantum Physics and Photonic Quantum Information, Ministry of Education, University of Electronic Science and Technology of China, Chengdu 611731, China

[4]Key Lab of Optical Fiber Sensing and Communication Networks, University of Electronic Science and Technology of China, Chengdu 611731, China

[5]Southwest Institute of Technical Physics, Chengdu 610041, China

[6]State Key Laboratory of Materials for Integrated Circuits, Shanghai Institute of Microsystem and Information Technology, Chinese Academy of Sciences, Shanghai 200050, China

[7]Institute of Systems Engineering, AMS, Beijing 100141, China

[8]CAS Center for Excellence in Quantum Information and Quantum Physics, University of Science and Technology of China, Hefei 230026, China

Corresponding author. Email: [†]zhouheng@uestc.edu.cn; [#]guokai07203@hotmail.com (KG); [*]zhouqiang@uestc.edu.cn (QZ)

**Abstract**

Dissipative Kerr solitons in optical microresonators underpin chip-scale frequency combs with applications ranging from coherent telecommunications to precision spectroscopy. Yet the characterization of their intrinsic femtosecond temporal structure remains challenging, as the low pulse energy and broad spectral bandwidth necessitate optical amplification and careful dispersion compensation in conventional ultrafast diagnostics, both of which can significantly distort the waveform. Here we demonstrate a quantum-interference metrology of microcomb solitons based on Hong-Ou-Mandel interference. By attenuating the soliton stream to the single-photon level and measuring fourth-order interference, we directly retrieve near transform-limited pulse durations without amplification or dispersion management, remaining accurate even after propagation through 25 km of standard fiber. The same interferogram also provides direct access to the temporal separations in multi-soliton states by converting inter-soliton separations into additional interference dips at corresponding delays, enabling sub-picosecond characterization of their intracavity temporal structure. This quantum-inspired paradigm introduces a fundamentally new metrological approach that is immune to amplification and dispersion distortions, offering a powerful tool for the characterization of complex soliton physics.

## Introduction

Optical frequency combs provide a phase-coherent bridge between the radio-frequency and optical domains and have become central to precision metrology and spectroscopy *(1–4)*. Kerr frequency combs generated in high-Q microresonators, specifically dissipative Kerr solitons (DKSs), offer ultrabroad spectra, low noise, microwave-to-terahertz mode spacings, and native compatibility with integrated photonic circuits*(5–12)*. These properties have enabled breakthroughs *(8, 9, 13, 14)* ranging from wavelength-division-multiplexed communications*(15)*, laser metrology*(16, 17)*, and spectroscopy*(18)*, to chip-scale optical clocks *(19)* and frequency synthesizers*(20)*. Beyond these classical applications, microcombs have also attracted attention in quantum photonics, where they have been employed for entangled photon-pair generation, quantum interference, and scalable quantum networking*(21–23)*.

Despite this progress, measuring the intrinsic femtosecond waveform of microcomb solitons remains challenging because the pulse energy per round trip is typically at the picojoule level while the spectral bandwidth extends over tens of terahertz. Conventional nonlinear techniques, such as frequency-resolved optical gating (FROG) and spectral phase interferometry for direct electric-field reconstruction (SPIDER)*(24–26)*, require high peak powers and thus rely on optical amplification. Such amplification is commonly enabled by chirped pulse amplification (CPA), which temporally stretches and subsequently recompresses pulses to mitigate nonlinear distortions*(27, 28)*. Linear alternatives, such as temporal imaging based on time lenses *(29–31)* and coherent optical sampling*(32–34)*, offer significantly enhanced sensitivity but typically demand stringent synchronization between the signal and the reference/pump source. Crucially, both classes of techniques are highly susceptible to group velocity dispersion (GVD) in the delivery fibers. Propagation of broadband pulses through even short fiber lengths induces significant chirp, obscuring the intrinsic transform-limited character of the solitons. Recovering the original waveform thus requires rigorous dispersion precompensation or model-dependent deconvolution based on a priori assumptions, reducing robustness and reliability, particularly for complex and nonrepetitive multi-soliton states. These constraints motivate an amplification-free and dispersion-immune metrology that enables the extraction of the intrinsic pulse width and the characterization of intra-cavity temporal separations.

Hong-Ou-Mandel (HOM) interference was originally developed as a probe of photon indistinguishability*(35)*, but its sensitivity to the temporal and spectral overlap of optical wave packets also makes it a powerful tool for ultrafast metrology at the single-photon level*(36)*. Unlike classical autocorrelation, which measures convolutions of intensity envelopes, HOM interference arises from fourth-order coherence and directly probes the overlap of field amplitudes. The overlap is insensitive to common spectral phase distortions, HOM interference is robust against dispersion*(37, 38)*. These properties suggest that HOM interference can be repurposed from a test of quantum indistinguishability into a metrological tool for characterizing ultrafast optical wave packets in regimes where classical nonlinear diagnostics are impractical.

Here, we demonstrate an amplification-free and dispersion-immune approach to soliton characterization by harnessing the physics of HOM interference. By attenuating the solitons to the single-photon level and measuring the fourth-order coherence function, we extract an intrinsic soliton pulse width of 83.4±2.1 fs from the interference dip without optical amplification or dispersion compensation, which agrees with the Fourier transform limit inferred from the optical spectrum. The retrieved width remains unchanged after propagation through a 25-km length of standard single-mode fiber, demonstrating an intrinsic robustness to group-velocity dispersion. The same HOM interferogram provides a direct temporal map of multi-soliton states, enabling sub-picosecond resolution over a cavity round trip. These results open a new avenue for bridging the gap between nonlinear and quantum optics, providing an essential diagnostic tool for understanding soliton physics in both classical and quantum technologies.

## Results

### Theoretical model and analysis.

A single DKS circulating in a microresonator is well approximated by a transform-limited sech pulse*(13)*,

$$E(t) = \frac{1}{\sqrt{2T_0}} \operatorname{sech}\left(\frac{t}{T_0}\right) \tag{1}$$

where $T_0$ determines the soliton duration. In a HOM interferometer, two identical wavepackets incident on a 50:50 beamsplitter with a temporal delay $\tau$ yield a coincidence probability

$$P_c(\tau) = \frac{1}{2}[1 - V|g(\tau)|^2], \tag{2}$$

where $g(\tau) = \int E(t)E^*(t-\tau)\,dt$ is the temporal overlap function, and $V$ is the visibility of HOM interference curve. For attenuated coherent states, multiphoton components limit the raw HOM visibility to 50%. Deviations from this limit primarily arise from distinguishability, such as polarization and spectral/temporal mismatch.

For the sech soliton waveform, this overlap takes the closed analytical form

$$g(\tau) = \frac{\tau/\mathrm{T}_0}{\sinh(\tau/\mathrm{T}_0)}. \tag{3}$$

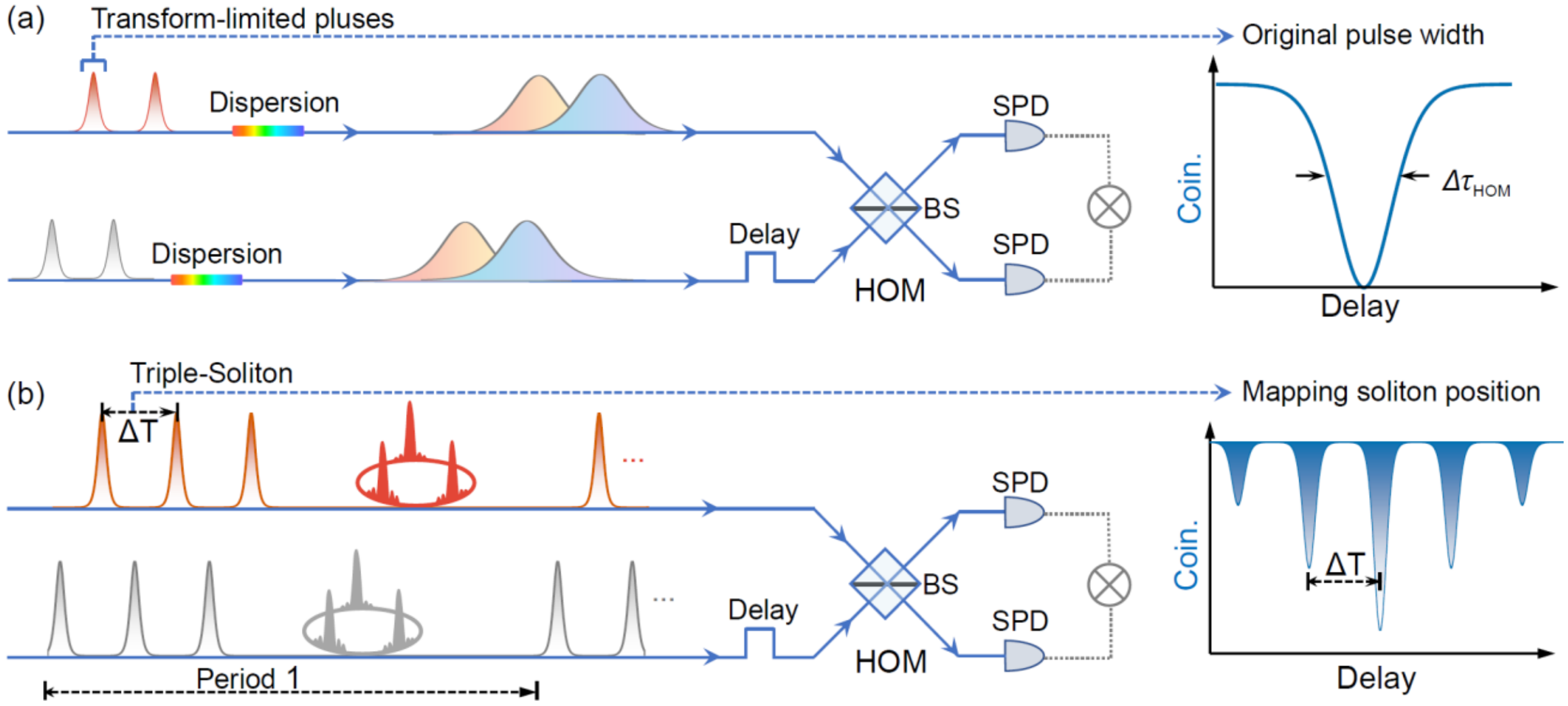

**Fig. 1. Conceptual illustration of quantum-interference metrology of dissipative Kerr solitons. (a)** Retrieve transform-limited pulse widths without amplification or dispersion compensation. **(b)** Characterization of the fine temporal structure of triple-soliton states by mapping their relative delays onto the interference dip positions.

See more details in Supplementary Materials Note1. The HOM dip width is therefore $\Delta\tau_{\mathrm{HOM}} = 2.98\,T_0$, which is independent of dispersion. Besides, the soliton full width at half maximum (FWHM) is $\Delta t_{\mathrm{FWHM}} = 1.76\,T_0$. This gives a direct conversion between the measured HOM dip width and the intrinsic soliton duration:

$$\Delta t_{\mathrm{FWHM}} = \Delta\tau_{\mathrm{HOM}}/1.69, \tag{4}$$

providing a dispersion-insensitive route to recovering the transform-limited pulse width. Because the HOM interference depends on the symmetry of the two-photon wave function rather than on the classical intensity profile, the retrieved width is insensitive to dispersion, allowing the intrinsic soliton duration to be recovered after propagation through the same dispersion environment as shown in Fig. 1(a)*(37)*.

We consider a multi-soliton state consisting of $M$ identical pulses at temporal positions $\{t_m\}$,

$$E_M(t) = \frac{1}{\sqrt{M}} \sum_{m=1}^{M} E(t - t_m). \tag{5}$$

Its HOM interference is governed by the multi-soliton overlap function

$$g_M(\tau) = \frac{1}{M} \sum_{m,n} g[\tau - (t_m - t_n)]. \tag{6}$$

Substitution into Eq. 2 reveals that HOM trace contains, in addition to the central dip at $\tau = 0$, a set of dips appearing at delays $\Delta T = t_m - t_n$, each corresponding to a pairwise soliton separation as shown in Fig. 1(b). The widths of all dips remain determined by $T_0$, enabling simultaneous extraction of both the soliton temporal width and the multi-soliton distribution directly from a single HOM measurement. See more details in Supplementary Materials.

**Temporal characterization of single-soliton microcombs**

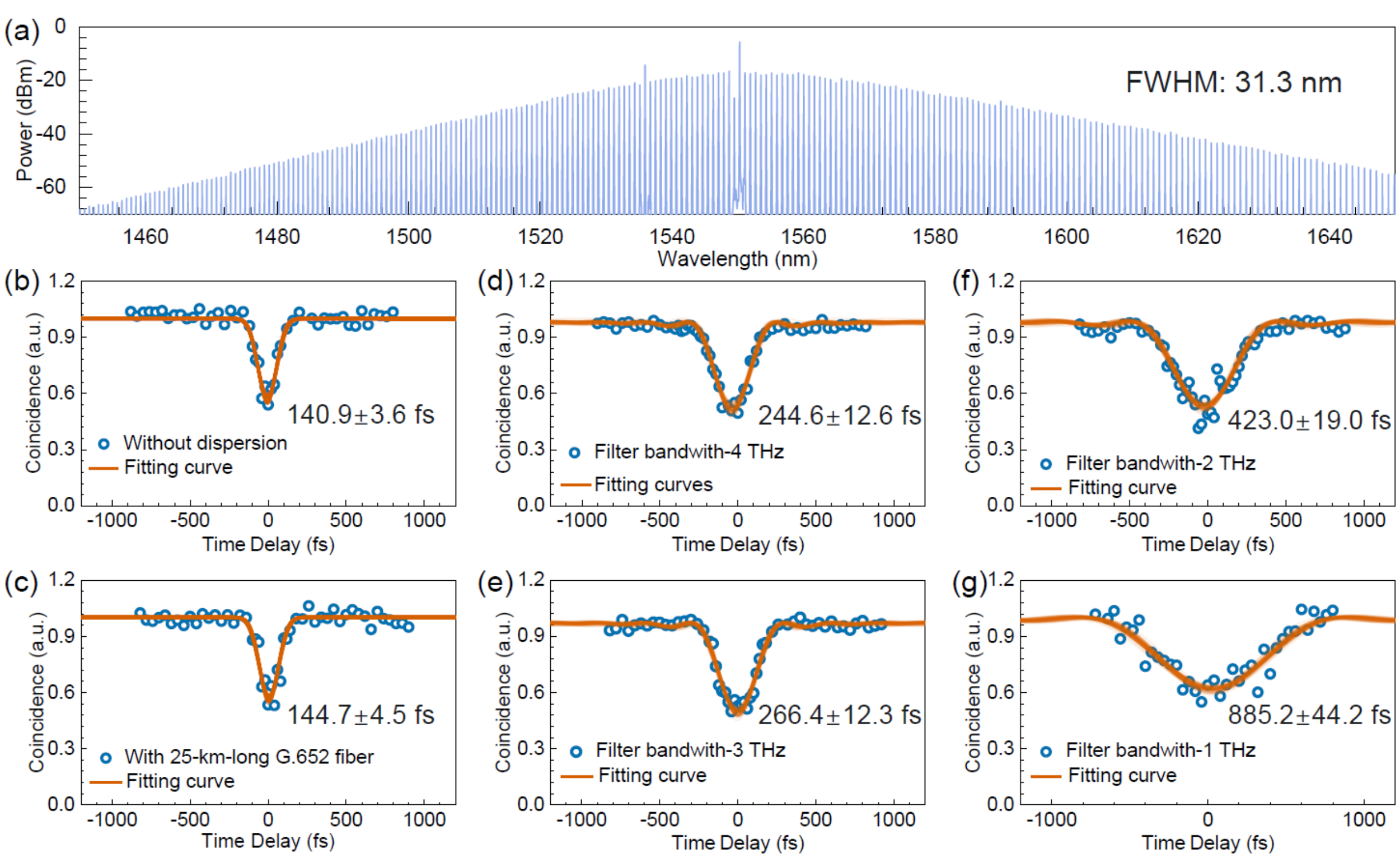

**Fig. 2. HOM-based temporal characterization of single-soliton microcombs. (a)** Optical spectrum of the single-soliton state exhibiting a $\mathrm{sech}^2$ envelope with a FWHM of 31.3 nm, corresponding to a Fourier-transform-limited duration of 80.6 fs. **(b)** HOM interference without additional dispersion, showing a dip width of $140.9 \pm 3.6$ fs and a raw visibility of $42.94 \pm 0.94\%$, yielding an intrinsic soliton duration of $83.4 \pm 2.1$ fs. **(c)** HOM interference after transmission through 25 km of G.652 fiber, demonstrating a nearly unchanged dip width of $144.7 \pm 4.5$ fs and hence dispersion-immune temporal retrieval. **(d)-(g)** HOM interference under spectral filtering with bandwidths of 4, 3, 2, and 1 THz, respectively. The dip broadens from $244.6 \pm 12.6$ fs to $885.2 \pm 44.2$ fs as the bandwidth decreases, while the visibility remains near the weak-coherent-state limit of $50\%$.

The DKSs are generated via auxiliary laser heating *(12)* See more details in Supplementary Materials. Figure 2(a) shows the single-soliton spectrum with a $\mathrm{sech}^2$ envelope with an FWHM of 31.3 nm, corresponding to a Fourier-transform-limited duration of 80.6 fs. The soliton stream is first split by a 50:50 fiber coupler and injected into the two arms of the interferometer. Indistinguishability in the polarization degree of freedom is ensured using polarizing beam splitters, while temporal indistinguishability is established by matching the optical path lengths with a variable delay line. Attenuators are inserted in both arms to equalize the mean photon number in the two paths at the single-photon level. The two outputs of the interferometer are detected by single-photon detectors, which convert the optical signals into electrical pulses that are processed by a time-to-digital converter to record coincidence events.

The measured coincidence counts as a function of relative delay are shown in Fig. 2(b). In the absence of additional dispersion, we observe a pronounced HOM dip with an FWHM of $140.9 \pm 3.6$ fs and a raw visibility of $42.94\% \pm 0.94\%$. With Eq. 4, we extract an intrinsic soliton temporal width of $83.4 \pm 2.1$, in excellent agreement with the Fourier-transform limit of 80.6 fs. To test dispersion immunity, we insert 25 km of ITU-T G.652 fiber outside the HOM interferometer, i.e., between BS1 and BS2 as shown in Fig. S1. Although the classical intensity waveform is broadened, the HOM dip in Fig. 2(c) remains unchanged - $144.7 \pm 4.5$ fs, confirming that the inferred intrinsic soliton width is insensitive to fiber-induced dispersion.

We further examine the effect of spectral filtering, which is commonly required in conventional temporal characterization techniques. In our experiment, the selected comb spectrum can be approximated by a rectangular profile, resulting in a sinc$^2$-shaped temporal intensity distribution. Correspondingly, the HOM interference envelope also follows a sinc$^2$-shaped profile such that the FWHM of the HOM dip is equal to the temporal FWHM of the single-photon wavepacket. More details are provided in the Supplementary Materials. Figures 2(d)–(g) show the HOM interference curves with filtering bandwidths of 4, 3, 2, and 1 THz, yielding FWHM of 244.6±12.6 fs, 266.4±12.3 fs, 423.0±19.0 fs, and 885.2±44.2 fs, respectively. These values are in good agreement with the corresponding pulse widths of 221.5 fs, 295.3 fs, 443.0 fs, and 886.0 fs derived from the Fourier transforms of the filtered spectra, further confirming the validity of the HOM-based temporal characterization under spectral filtering.

**Temporal Mapping of Multi-Soliton States.**

Beyond single-pulse envelope characterization, HOM interference provides direct access to the intracavity temporal structure of multi-soliton states. The micro-ring resonator used in our experiment has a free spectral range (FSR) of 100 GHz, corresponding to a cavity round-trip time of $T_R = 10\ ps$. By scanning the relative delay from -5 ps to 15 ps, we map the relative temporal positions of solitons within a round trip. Figures 3(a) and (b) show the optical spectrum and HOM interference for a single-soliton state. Coincidence dips appear at delays of 0 ps and 10 ps, reflecting the cavity repetition period and the high temporal coherence between successive pulses.

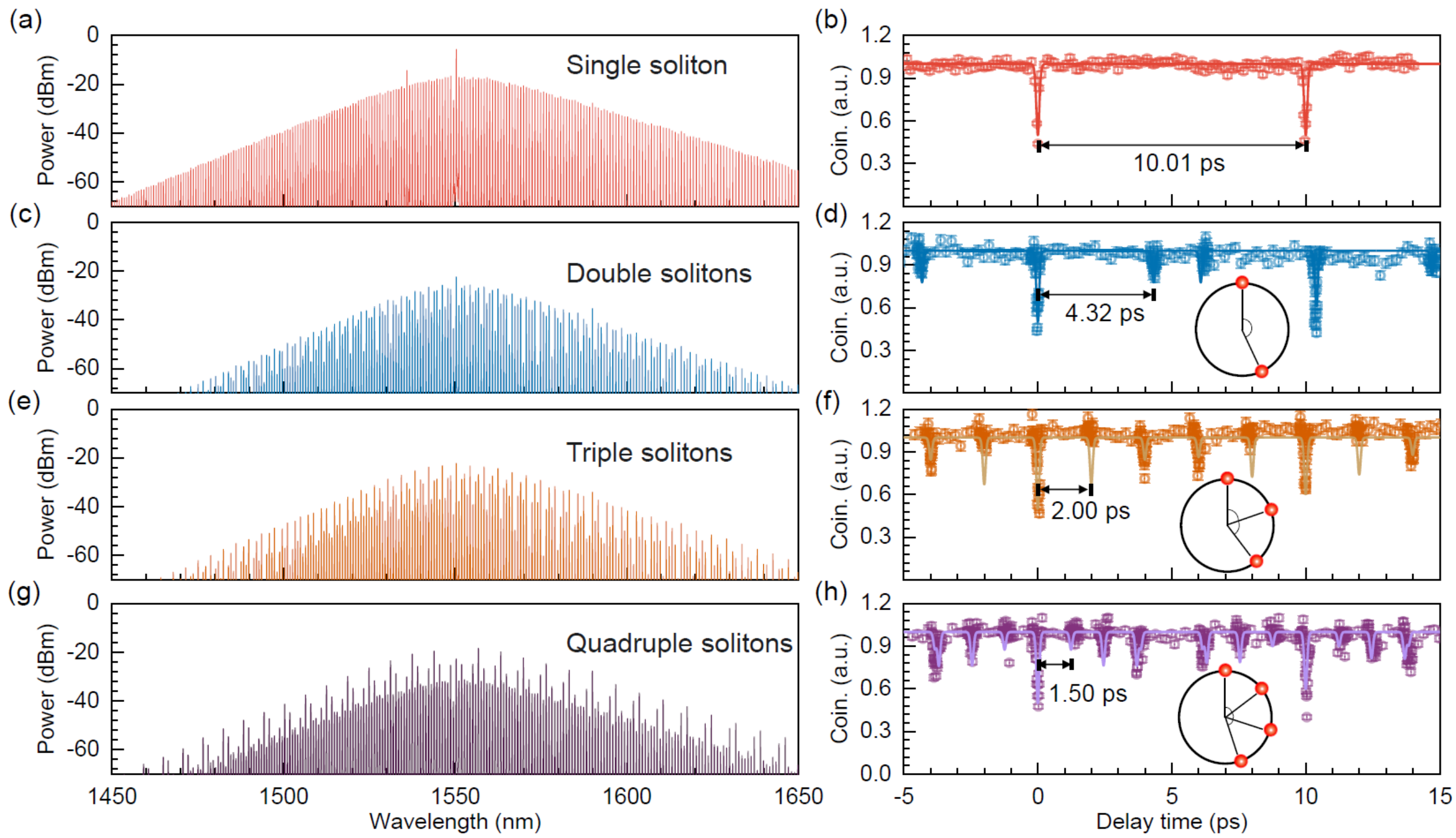

**Fig. 3. Temporal mapping of multi-soliton states via HOM interference. (a, c, e, g)** Optical spectra of single-, double-, triple-, and quadruple-soliton states, respectively. **(b, d, f, h)** Corresponding HOM interference traces. The free spectral range of 100 GHz corresponds to a temporal periodicity of 10 ps. The extracted inter-soliton separations are 4.3 ps (double soliton), 2.0 ps (triple soliton), and 1.5 ps (quadruple soliton). Open symbols denote experimental data and solid lines are multi-peak fits. The insets in (d, f, h) illustrate the relative circulating positions of the solitons within the micro-ring cavity.

As the pump detuning is tuned to trap multiple solitons, the HOM interference evolves to reveal their relative temporal arrangement. For a double-soliton state, as shown in Figs. 3(c) and (d), the interference exhibits dips at 4.32 ps (and the complementary 5.68 ps within the same round trip), indicating a stable state with a fixed inter-soliton separation. See more details in Supplementary Materials for RF characterization of phaser-locked Kerr soliton states. For triple- and quadruple-soliton states, the interference patterns become richer while remaining clearly resolvable. As shown in Figs. 3(f) and (h), periodic dips with spacings of 2.00 ps and 1.50 ps are observed, respectively. These results are consistent with the formation of solitons, where pulses are arranged with near-equidistant spacing. The insets in Figs. 3(d, f, h) show the relative soliton positions in the micro-ring cavity. The ability to resolve sub-round-trip temporal separations without ultrafast photodetectors or classical cross-correlation highlights the sensitivity of quantum interference. This HOM-based temporal mapping provides a robust and power-efficient probe of soliton binding and crystallization dynamics in Kerr microcombs.

## Discussion

We have demonstrated that HOM interference provides a robust and precise probe of DKSs in microresonators at the single-photon level. Unlike classical temporal diagnostics based on high-order nonlinearities, the HOM approach probes the indistinguishability of wave packets and is therefore sensitive to their intrinsic temporal structure. Our approach offers three distinct advantages over conventional ultrafast characterization techniques. First, operating without optical amplification, the HOM measurement provides a minimally invasive probe of intracavity dynamics and enables direct temporal mapping of multi-soliton states with sub-picosecond resolution. This allows the relative temporal spacings of soliton molecules and crystals to be characterized without ultrafast detectors or classical cross-correlation. Second, a central result is the demonstrated insensitivity to group-velocity dispersion. Whereas classical autocorrelation measures convolutions of intensity envelopes and is strongly distorted by dispersion, HOM interference depends on wave-packet overlap. As a result, the intrinsic soliton duration is accurately retrieved even after transmission through 25 km of dispersive fiber, indicating that the coherence and modal structure of the microcomb are preserved at the single-photon level. It should be noted, for an unchirped pulse, such as an ideal dissipative Kerr soliton, the duration extracted from the HOM measurement directly corresponds to the transform-limited pulse width and therefore accurately reflects the intrinsic temporal width of the pulse. In contrast, for a chirped pulse, the HOM measurement generally retrieves the transform-limited temporal scale rather than the actual temporal waveform. Third, while inter-soliton separations are in principle encoded as spectral modulation, extracting them from an intensity-only spectrum is ambiguous because spectral phase is lost and fine fringes can be washed out by finite resolution and noise *(39)*. In contrast, the HOM interferogram maps each pairwise separation directly onto the delay axis as interference dips, providing a direct measurement of the inter-soliton separations that is decoupled from spectral-resolution limits. Furthermore, our method can be further employed to characterize higher-order multi-soliton states with different temporal configurations *(40)*. Despite these advantages, our current implementation becomes challenging for densely packed soliton states. For instance, in our present experiments, the maximum resolvable soliton number is estimated to be approximately 8 based on the statistical fluctuations of the coincidence counts. As the soliton number increases, the weakest side dips become difficult to distinguish from statistical noise, while overlap between neighboring HOM dips further limits the maximum resolvable soliton number. Nevertheless, this limitation is not fundamental and can be alleviated by reducing the statistical uncertainty of the coincidence, for example by increasing the single-photon count rates, extending the integration time, or increasing the coincidence window.

In summary, we have proposed and demonstrated an amplification-free and dispersion-immune approach to soliton characterization based on HOM interference. A pulse duration of $83.4 \pm 2.1$ fs is extracted from the interference dip without optical amplification or dispersion compensation, consistent with the Fourier-transform limit. After propagation through 25 km of standard single-mode fiber, the intrinsic pulse duration is still accurately retrieved. The temporal structure of multi-soliton states is also characterized, with inter-soliton spacings of 4.32, 2.00, and 1.50 ps for double-, triple-, and quadruple-soliton states, respectively. Our work establishes a direct link between quantum interference and ultrafast nonlinear dynamics, illustrating a form of quantum-enhanced classical metrology. The combination of dispersion robustness, single-photon sensitivity, and high temporal resolution makes this approach applicable in regimes where conventional high-power techniques are impractical, and provides a new tool for studying nonequilibrium dynamics and emergent order in driven dissipative photonic systems.

## Materials and Methods

### Experimental setup

Figure S1 shows our experimental setup. A $Si_3N_4$ microring resonator (MRR) is utilized for the generation of Kerr optical frequency comb. The microring is designed with a width-height cross-section of $1600 \times 800$ nm$^2$ for proper anomalous dispersion (the TE mode is adopted). The loaded Q-factor of ~$10^6$ with free spectral ranges (FSRs) of around 100 GHz. The $Si_3N_4$ chips are packaged with polarization-maintaining I/O fibers for easy and stable external driving with an insertion loss of 5.0 dB. In our experiments, a narrow-linewidth fiber laser is launched into the MRR. The polarization state of the pump laser is manipulated by a polarization controller (PC). An Er-doped fiber amplifier (EDFA) is adopted to boost the pump power. The sideband noise of the EDFA is cleaned by a dense-wavelength division multiplexer (DWDM) with an extinction ratio of $>$ 40 dB. To suppress the unwanted cavity thermal dynamics during soliton generation, the method of auxiliary laser heating is adopted. An auxiliary laser from another fiber laser at a wavelength of 1535.82 nm is amplified and launched into the microcavity from the opposite direction to the pump laser. As the auxiliary lasers enter the cavity mode from the blue detuning range, the micro-resonator is heated, and all the resonances are thermally redshifted. By properly setting the frequency and power of the auxiliary laser, the heat flow caused by the pump and auxiliary lasers can be well balanced out, making the pump laser stably scan across the entire resonance and enter the red-detuned soliton existence region without notable thermal dragging. In our experiments, the wavelengths of the pump and auxiliary laser are 1550.08 nm and 1535.82 nm with powers of 430 mW and 750 mW, respectively. The generated solitons from Cir2 are sent to two cascaded DWDMs at ITU channels of C34 and C52 to suppress the residual pump and auxiliary light. The bandwidth of the DWDMs is ~0.6 nm with a channel spacing of 100 GHz and an isolation of ~40 dB. The optical powers of the single, double, triple, and quadruple solitons monitored by the power meter (PM) are -13.6, -10.6, -9.6, and -8.2 dBm, respectively.

To characterize the Kerr optical frequency comb, we use the HOM interferometer to measure the temporal width of the ultrashort Kerr soliton pulse as shown in the middle of Fig. S1. A 50/50 BS is used to separate the attenuated laser pulses into two paths. Two wave-packets are injected into a HOM interferometer, which consists of two variable optical attenuators (VOA), two polarization beam splitters (PBS), a 50/50 polarization-maintaining BS, and a fiber-pigtailed variable optical delay-line (MDL). Two VOAs in two arms are used to adjust the mean photon number in the two optical paths, which ensures that the mean photon numbers of the two paths are the same. Two PBSs are used to ensure indistinguishability in polarization. The relative time delay is introduced by an optical delay line (MDL) with an accuracy of 10 fs and a maximum range of 560 ps. Two output ports of the HOM interferometer are connected with two superconducting nanowire single-photon detectors (SNSPDs). The electronic signals generated by photon detection events are input into a time-to-digital converter (TDC) to obtain the coincidence counts. The HOM dip can be observed in the HOM interference curve, i.e., coincidence counts versus the relative time delay between two paths. In our experiments, the detected photon count rates are set as 400 kHz. The coincidence counts are approximately 17000 outside the HOM dip and decreased to approximately 8800 at the bottom of the dip with a coincidence window of 10 ns and an integration time of 10 s. The detection efficiency of the SNSPDs is ~85% with a dark count rate of ~50 Hz and a timing jitter of ~50 ps.

## Acknowledgments

**Funding:** This work was supported by Quantum Science and Technology-National Science and Technology Major Project (Nos. 2024ZD0300800, 2021ZD0300701), National Natural Science Foundation of China (Nos. 62475039, 62405046, 62375043), Sichuan Science and Technology Program (Nos. 2024YFHZ0369, 2024YFHZ0370, 2024YFHZ0368), Tianfu Jiangxi Laboratory (No. TFJX-ZD-2025-005).

**Author contributions:**

Conceptualization: HZ, KG, QZ
Methodology: YRF, YG, JL, YJH, HZS, HL, LXY, KQ
Investigation: YRF, YG
Visualization: YRF, YG, JL, YJH, HZS, HL, LXY, KQ, QZ
Supervision: HZ, KG, GCG, QZ
Writing—original draft: YRF, QZ
Writing—review & editing: YRF, YG, HZ, KG, QZ

**Competing interests:** Authors declare that they have no competing interests.

**Data and materials availability:** All data needed to evaluate the conclusions in the paper are present in the paper and/or the Supplementary Materials. Additional data related to this paper may be requested from the authors.